\documentclass[prb,aps,twocolumn,showpacs,showkeys]{revtex4}

\usepackage{epsfig}
\usepackage{amsmath,amssymb,amsfonts}
\usepackage{bm}

\begin{document}

\title{Formation of long-range spin distortions by a bound
magnetic polaron}

\author{S.L.~Ogarkov, M.Yu.~Kagan}
\affiliation{Kapitza Institute for Physical Problems, Russian
Academy of Sciences, Kosygina str. 2, Moscow, 119334 Russia}

\author{A.O.~Sboychakov, A.L.~Rakhmanov, and K.I.~Kugel}
\affiliation{Institute for Theoretical and Applied
Electrodynamics, Russian Academy of Sciences, Izhorskaya str.
13/19, Moscow, 125412 Russia}

\begin{abstract}
The structure of bound magnetic polarons in an antiferromagnetic
matrix is studied in the framework of two-dimensional (2D) and
three-dimensional (3D) Kondo-lattice models in the double exchange
limit ($J_H\gg t$). The conduction electron is bound by a
nonmagnetic donor impurity and forms a ferromagnetic core of the
size about the electron localization length (bound magnetic
polaron). We find that the magnetic polaron produces rather
long-range extended spin distortions of the antiferromagnetic
background around the core. In a wide range of distances, these
distortions decay as $1/r^2$ and $1/r^4$ in 2D and 3D cases,
respectively. In addition, the magnetization of the core is
smaller than its saturation value. Such a magnetic polaron state
is favorable in energy in comparison to usually considered one
(saturated core without extended distortions).
\end{abstract}



\date{\today}

\maketitle

\section{Introduction}

The physics of electronic phase separation is very popular
nowadays in connection with the strongly correlated electron
systems, especially, in materials with the colossal
magnetoresistance such as manganites~\cite{dagbook,DagSci}. The
phase separation manifests itself in other magnetic materials such
as cobaltites~\cite{cobalt}, nickelates~\cite{nickel} and also in
low-dimensional magnets~\cite{low-d}. The nanoscale
inhomogeneities were also widely discussed for heavy fermion
compounds~\cite{fermion}, string oscillators in magnetic
semiconductors~\cite{Bul}, electron-hole droplets in
semiconducting nanostructures~\cite{Keld}, and high-Tc
superconductors~\cite{Zaanen,Castel,NagSup}, in particular, in
relation to the phase separation in the $t-J$ model~\cite{Emery}
and to the formation of paramagnetic spin bags~\cite{Schrieffer}.

The formation of small ferromagnetic (FM) metallic droplets
(magnetic polarons or ferrons) in antiferromagnetic (AFM)
insulating matrix was discussed beginning from works
Refs.~\onlinecite{Nag67,Kasuya}. In application to manganites,
this problem was addressed in a number of papers (see for the
review Ref.~\onlinecite{dagbook}). Usually in manganites, the
'rigid' (well-defined) magnetic polarons with very rapidly
decreasing tails of magnetic distortion are considered. In other
words, the intermediate region where the canting angle $\nu$
changes from $0$ (FM domain) to $\pi$ (AFM domain) is narrow (of
the order of interatomic distance $d$), so the radius of the
magnetic polaron $a$ is a well defined quantity. Moreover, the
tails of magnetic distortions exponentially decrease for the
distances $r>a$ (outside the ferron).

Long ago, de Gennes discussed the possibility of a slow decay of
the AFM order distortions due to the magneto-dipole
interaction~\cite{DeGen}. An attempt to get such a type of
'coated' magnetic polarons was made for a one-dimensional (1D) AFM
chain in Refs.~\onlinecite{Nag01b,Ivan1}. These calculations show
that the characteristic length of the distorted spin surrounding
can be much larger than the size of the trapping region. However,
in 1D, the 'coated' ferrons turn out to be metastable objects
while 'rigid' magnetic polarons correspond to the ground
state~\cite{Fer1D}. A continuous 3D model of 'coated' magnetic
polarons with the account taken of the tails of electron wave
function in the AFM matrix (which appeared when we consider
quantum nature of canting) was studied in Ref.~\onlinecite{Klap}.
Unfortunately, even in this more advanced approach, the tails of
the magnetic distortions exponentially decrease outside the
ferron. However, in Ref.~\onlinecite{Klap}, the gradient
contribution to the magnetic energy proportional to $(\nabla
\nu)^2$ was not included to the variational procedure to reduce
the order of the differential equations under study. In the
present paper, we generalize the 1D model considered in
Ref.~\onlinecite{Fer1D} to the 2D and 3D cases. In contrast to the
approach of Ref.~\onlinecite{Klap}, we take into account the
magnetic anisotropy, discrete nature of the lattice, and the
gradient terms. Appropriate account of the gradient terms leads to
a slower decrease of magnetic distortions outside the ferron.

In our model, we consider an electron, which is bound at a donor
impurity by the Coulomb attractive potential. The Coulomb
potential $V$ is assumed to be strong in comparison to the other
relevant interactions. Namely $V\sim J_H \gg t \gg
J_{\text{dd}},K$, where $J_H$ is the Hund's rule coupling,
$J_{\text{dd}}$ is the AFM exchange interaction, $t$ is a hopping
integral, $K$ is a constant of the magnetic anisotropy. In this
range of parameters, the radius of the electron localization is of
the order of interatomic distance $d$. The magnetic distortions
outside the electron localization region (ferron core) are shown
to decrease with the distance $r$ as $1/r^4$ and $1/r^2$ in the 3D
and 2D cases, respectively. In contrast to the 1D
model~\cite{Fer1D}, these 'coated' ferrons correspond to the
ground state of the system.

Note that the 2D case widely discussed in our paper could be
rather relevant to experiment since a lot of magnetic oxides
including manganites and cobaltites have layered magnetic
structure and a 2D structure is simply a limiting case of the
layered system. Actually, as it was shown by neutron diffraction
studies~\cite{HennPRL98,HennPRB00,HennPRL05, HennNJPh05}, the
ferrons in low-doped manganites could be 'platelets'
characteristic for 2D rather than spherical droplets
characteristic for 3D.

At the end of the paper, we find the applicability range of the
proposed approach by taking into account the tails of the electron
wave function outside the ferron. Our results enrich the concept
of microscopic phase separation for the localized magnetic
polarons.

\section{The model Hamiltonian}

We consider 2D or 3D cubic lattice of antiferromagnetically
coupled local spins $\mathbf{S_n}$ treated as classical vectors.
We assume that the crystal has uniaxial magnetic anisotropy, with
$x$ axis being the easy axis. The nonmagnetic donor impurities are
placed in the center of some unit cells of the lattice. It is
assumed that the doping concentration is small enough and
therefore we can consider an isolated impurity and restrict
ourselves to a single-electron problem. The Hamiltonian of such a
system has the following form:
\begin{eqnarray}
\hat{H}&=&\hat{H}_{\text{el}}+J_{\text{dd}}\sum_{\langle\mathbf{nm}\rangle}%
\left[\mathbf{S_nS_m}+S^2\right]\label{H}\\%
&&-K'\sum_{\mathbf{n}}\left[\left(S_{\mathbf{n}}^{x}\right)^2-S^2\right],\nonumber\\
\hat{H}_{\text{el}}&=&-t\sum_{\langle\mathbf{nm}\rangle\sigma}%
\hat{a}^{\dag}_{\mathbf{n}\sigma}\hat{a}_{\mathbf{m}\sigma}%
-\frac12J_H\sum_{\mathbf{n}\sigma\sigma'}\hat{a}^{\dag}_{\mathbf{n}\sigma}%
\left(\mathbf{S_n}{\bm\sigma}\right)_{\sigma\sigma'}\hat{a}_{\mathbf{n}\sigma'}\label{Hel}\nonumber\\%
&&-V\sum_{\mathbf{n}\sigma}\frac{\hat{a}^{\dag}_{\mathbf{n}\sigma}\hat{a}_{\mathbf{n}\sigma}}%
{\left|\mathbf{n-n}_0\right|},
\end{eqnarray}
where $\hat{a}^{\dag}_{\mathbf{n}\sigma}$,
$\hat{a}_{\mathbf{n}\sigma}$ are the creation and annihilation
operators for the conduction electron with spin projection
$\sigma$ at site $\mathbf{n}$, ${\bm\sigma}$ are Pauli matrices,
and the symbol $\langle\dots\rangle$ denotes the summation over
nearest neighbors. The second and third terms in Eq.~\eqref{H} are
the AFM exchange between local spins and the magnetic anisotropy
energy, respectively. The first two terms in $\hat{H}_{\text{el}}$
describe the kinetic energy of conduction electrons, and the
Hund's rule coupling between the conduction electrons and the
localized spins. The last term in Eq.~\eqref{Hel} describes the
Coulomb interaction between electrons and the impurity ion placed
in the middle of the main diagonal of the unit cell (e.g., in the
3D case, between lattice cites $\mathbf{n}=\{0,\,0,\,0\}$ and
$\mathbf{n}=\{1,\,1,\,1\}$, that is,
$\mathbf{n}_0=\{1/2,\,1/2,\,1/2\}$).

Below we consider the range of parameters $J_H\gg t\gg
J_{\text{dd}}\gg K'$ characteristic of manganites. In the limit
$J_H\to\infty$, the spin of electron at site $\mathbf{n}$ is
parallel to the local spin
$\mathbf{S_n}=S\{\sin\theta_{\mathbf{n}}\cos\varphi_{\mathbf{n}},\,%
\sin\theta_{\mathbf{n}}\sin\varphi_{\mathbf{n}},\,\cos\theta_{\mathbf{n}}\}$.
This implies the transformation of $\hat{a}_{\mathbf{n}\sigma}$ to
operators $\hat{c}_{\mathbf{n}}$ with spin projection $+1/2$ onto
the direction of ${\bf S_n}$ (spinless fermions) and also to the
transformation of hopping amplitudes: $t\to
t\cos(\nu_{\mathbf{nm}}/2)e^{-i\omega_{\mathbf{nm}}}$, where
$\nu_{\mathbf{nm}}$ is the angle between $\mathbf{S_n}$ and
$\mathbf{S_m}$
$$
\cos\nu_{\mathbf{nm}}=\cos\theta_{\mathbf{n}}\cos\theta_{\mathbf{m}}+%
\sin\theta_{\mathbf{n}}\sin\theta_{\mathbf{m}}\cos(\varphi_{\mathbf{n}}-\varphi_{\mathbf{m}}),
$$
and
$$
\omega_{\mathbf{nm}}=\arg\left[\cos\frac{\theta_\mathbf{n}}{2}\cos\frac{\theta_{\mathbf{m}}}{2}+%
\sin\frac{\theta_\mathbf{n}}{2}\sin\frac{\theta_{\mathbf{m}}}{2}%
\text{e}^{i(\varphi_{\mathbf{n}}-\varphi_{\mathbf{m}})}\right]
$$
is the Berry phase~\cite{Berry,MulDag}.

At low doping, the electron is bound by the impurity electrostatic
potential. In the limit of strong electron-impurity coupling
$V\to\infty$, the electron wave function $\Psi_{\mathbf{n}}$ will
be nonzero only at sites nearest to the impurity. Let us focus
first on the 2D case. The three-dimensional case is considered in
the similar way. Supposing that $\Psi_{\mathbf{n}}\neq0$ only for
$\mathbf{n}_1=\{1,\,1\}$, $\mathbf{n}_2=\{1,\,0\}$,
$\mathbf{n}_3=\{0,\,0\}$, and $\mathbf{n}_4=\{0,\,1\}$, and
following the standard diagonalization procedure, we find for
smallest eigenvalue of the electron energy
\begin{widetext}
\begin{eqnarray}\label{Eel}
E_{\text{el}}&=&-\displaystyle\frac{t}{\sqrt{2}}\left[c_{12}^2+c_{23}^2+c_{34}^2+c_{41}^2%
\phantom{\sqrt{\sin^2\displaystyle\!\!\frac{\Delta\omega}{2}}}\right.\nonumber\\
&&\left.+\sqrt{\left[(c_{12}-c_{34})^2+(c_{23}+c_{41})^2\right]\left[(c_{12}+c_{34})^2+(c_{23}-c_{41})^2\right]%
-16c_{12}c_{23}c_{34}c_{41}\sin^2\displaystyle\!\!\frac{\Delta\omega}{2}}\right]^{1/2}\!\!\!\!\!\!,
\end{eqnarray}
\end{widetext}
where the constant term $-J_HS/2-V\sqrt{2}$ is omitted. In
expression~\eqref{Eel},
$c_{ij}=\cos(\nu_{\mathbf{n}_i\mathbf{n}_j}/2)$,
$\omega_{ij}=\omega_{\mathbf{n}_i\mathbf{n}_j}$, and
$\Delta\omega=\omega_{12}+\omega_{23}+\omega_{34}+\omega_{41}$.
Note that in the general case we have $\Delta\omega\neq0$. The
electron energy $E_{\text{el}}$ has a minimum when all spins
$\mathbf{S}_{\mathbf{n}_i}$ are parallel to each other. Thus, we
have a bound magnetic polaron state: a ferromagnetic core embedded
in the antiferromagnetic matrix.

\section{Magnetic structure in the 2D case}\label{Mstruct}

In this section, we calculate in detail the magnetic structure of
bound magnetic polarons for the 2D lattice. The problem can be
solved by minimization of the total energy with respect to angles
$\theta_{\mathbf{n}}$ and $\varphi_{\mathbf{n}}$. It is convenient
to make first the following transformation. In the absence of
magnetic polaron, the lattice consists of two magnetic sublattices
with antiparallel magnetizations. In one sublattice, we perform
the transformation of the angles
$\theta_{\mathbf{n}}\to\pi-\theta_{\mathbf{n}}$ and
$\varphi_{\mathbf{n}}\to\text{mod}(\pi+\varphi_{\mathbf{n}},2\pi)$.
As a result, an AFM order becomes FM, and vice versa. Such a
transformation allows us to work with continuously changing
orientation of spins outside the ferron core. The total energy
then reads
\begin{eqnarray}\label{Etot}
E&=&E_{\text{el}}+J\sum_{\langle\mathbf{nm}\rangle}\left[1-\cos\nu_{\mathbf{nm}}\right]\nonumber\\%
&&-K\sum_{\mathbf{n}}\left[\sin^2\theta_{\mathbf{n}}\cos^2\varphi_{\mathbf{n}}-1\right],
\end{eqnarray}
where $J=J_{\text{dd}}S^2$, and $K=K'S^2$. Note that after the
transformation, we should replace in Eq.~\eqref{Eel} $c_{ij}\to
s_{ij}=\sin(\nu_{\mathbf{n}_i\mathbf{n}_j}/2)$ and
$\Delta\omega\to-\Delta\omega$.

Let the lattice plane be $xy$ plane. In the range of parameters
under study, the trivial solution of the problem Eq.~\eqref{Etot}
corresponds to the case $\theta_{\mathbf{n}}=\pi/2$,
$\varphi_{\mathbf{n}}=0$ for $\mathbf{n}\neq\mathbf{n}_i$, and
$\varphi_{\mathbf{n}_1}=\varphi_{\mathbf{n}_3}=0$,
$\varphi_{\mathbf{n}_2}=\varphi_{\mathbf{n}_4}=\pi$. So, we have a
bound magnetic polaron with completely polarized spins embedded in
purely antiferromagnetic background. The total magnetic moment of
such a polaron is parallel to the easy axis. We refer this trivial
solution as a 'bare' magnetic polaron. Its energy is
$E_{\text{p}}^{0}=-2t+16J$.

\begin{figure}
\centerline{\includegraphics[width=0.48\textwidth]{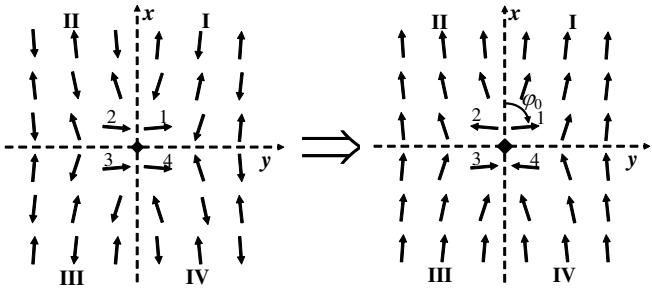}}
\caption{\label{FigCoatedFerron} 'Coated' magnetic polaron before
(left figure) and after (right figure) transformation of angles in
one sublattice. The magnetic structure is calculated by solving
Eqs.~\eqref{sys} at $t/J=50$, $\varkappa=5 \times 10^{-3}$. At
this values of parameters, $\varphi_0=85^{\circ}$.}
\end{figure}

There is another solution corresponding to the magnetic polaron
state with magnetic moment perpendicular to the easy axis: a
'coated' magnetic polaron. Let us assume again that all spins in
the lattice lie in the $xy$ plane, that is,
$\theta_{\mathbf{n}}=\pi/2$. It is clear from symmetry that angles
$\varphi_i\equiv\varphi_{\mathbf{n}_i}$ for the local spins inside
the magnetic polaron should satisfy the conditions
\begin{equation}\label{SymCond}
\varphi_1=\varphi_3=\varphi_0,\,\,\,\varphi_2=\varphi_4=-\varphi_0,
\end{equation}
where $0<\varphi_0\leq\pi/2$ (see Fig.~\ref{FigCoatedFerron}).
Minimizing the total energy Eq.~\eqref{Etot} with respect to
$\varphi_{\mathbf{n}}=\varphi_{n_x,n_y}$, taking into account
relation~\eqref{Eel} for $E_{\text{el}}$ and the fact that
$\Delta\omega=0$ for a planar configuration of spins, we find the
following set of nonlinear equations
\begin{eqnarray}\label{sys}
\sum_{{\bm\Delta}}\sin(\varphi_{\mathbf{n}+{\bm\Delta}}-\varphi_{\mathbf{n}})%
-\frac{\varkappa}{2}\sin2\varphi_{\mathbf{n}}=\nonumber\\%
=\frac{t}{2J}\sum_{i}\delta_{\mathbf{nn}_i}(-1)^{i}\cos\varphi_{i},
\end{eqnarray}
where $\varkappa=2K/J$, $\delta_{\mathbf{nm}}$ is the Kronecker
symbol, and ${\bm\Delta}$ takes values $\{\pm1,\,0\}$,
$\{0,\,\pm1\}$.

Equations~\eqref{sys} with conditions~\eqref{SymCond} are solved
numerically for the cluster containing $40\times40$ sites. The
further growth of the number of sites in the cluster does not
change the obtained results. The initial angle $\varphi_0$ is also
found. The calculated magnetic structure is shown in
Fig.~\ref{FigCoatedFerron}. We see from this figure that, in
contrast to the 'bare' ferron, the 'coated' magnetic polaron
produces spin distortions of AFM background outside the region of
electron localization. In addition, the 'coated' magnetic polaron
has a magnetization lower than its saturation value
($\varphi_0<\pi/2$). Moreover, the 'coat' has a magnetic moment
opposite to that of the core (sites 1--\,4 in Fig.~\ref
{FigCoatedFerron}).

In order to get analytical estimations for the spatial
distribution of the spin distortions, we find an approximate
solution to Eqs.~\eqref{sys} in the continuum limit. Namely,
angles $\varphi_{\mathbf{n}}$ are treated as values of continuous
function $\varphi(\mathbf{r})$ at points
$\mathbf{r}=\mathbf{n}-\mathbf{n}_0$. Assuming that outside the
magnetic polaron the following condition is met
$|\varphi_{\mathbf{n}+{\bm\Delta}}-\varphi_{\mathbf{n}}|\ll1$, we
can expand $\varphi(\mathbf{r}+{\bm\Delta})$ in the Taylor series
up to the second order in $\Delta$
\begin{equation}\label{Taylor}
\varphi(\mathbf{r}+{\bm\Delta})%
\approx\varphi({\mathbf{r}})+\Delta^{\alpha}\partial_{\alpha}\varphi({\mathbf{r}})+%
\frac12\Delta^{\alpha}\Delta^{\beta}\partial_{\alpha}\partial_{\beta}\varphi({\mathbf{r}})\,.
\end{equation}
Substituting this expansion into Eq.~\eqref{sys}, we find that
function $\varphi(\mathbf{r})$ outside the magnetic polaron should
satisfy the 2D sine-Gordon equation
\begin{equation}\label{SinGordon}
\Delta\varphi-\frac{\varkappa}{2}\sin2\varphi=0\,.
\end{equation}
In the range of parameters under study, $K\ll J$, that is
$\varkappa\ll1$, we can linearize this equation. As a result, we
obtain
\begin{equation}\label{LinEq}
\Delta\varphi-\varkappa\varphi=0\,.
\end{equation}
This equation should be solved with the boundary conditions at
infinity $\varphi(\mathbf{r})|_{r\to\infty}\!\!\to0$ and with some
boundary conditions at the surface of the magnetic polaron. We
model the magnetic polaron by a circle of radius $a=1/\sqrt{2}$
(in the units of lattice constant $d$) and choose the Dirichlet
boundary conditions
\begin{equation}\label{Dirichlet}
\varphi(\mathbf{r})|_{r=a}=\tilde{\varphi}(\zeta)\,,
\end{equation}
where we introduce polar coordinates $(r,\zeta)$ in the $xy$
plane. The function $\tilde{\varphi}(\zeta)$ can be found in the
following way. Note, that $\tilde{\varphi}(\zeta)$ should satisfy
the symmetry conditions~\eqref{SymCond} at points
$\zeta_i=\pi(2i-1)/4$
\begin{equation}\label{SymCond2}
\tilde{\varphi}\left(\zeta_i\right)=\varphi_i\,,\,\,i=1\dots4\,.
\end{equation}
Since the function $\tilde{\varphi}(\zeta)$ is a periodic one, it
can be expanded in the Fourier series
\begin{equation}\label{FS}
\tilde{\varphi}(\zeta)=\sum\limits_{m=0}^{\infty}\left[a_m\cos
m\zeta+b_m\sin m\zeta\right]\,.
\end{equation}
In Eq.~\eqref{FS} we neglect the terms with $m>2$ which allows us
to keep the minimum number of terms to satisfy the conditions
\eqref{SymCond2}. It follows from Eq.~\eqref{SymCond2} that
$a_0=a_1=a_2=b_1=0$. Finally, we obtain
\begin{equation}
\tilde{\varphi}(\zeta)=\varphi_0\sin2\zeta\,.
\end{equation}

\begin{figure}
\centerline{\includegraphics[width=0.4\textwidth]{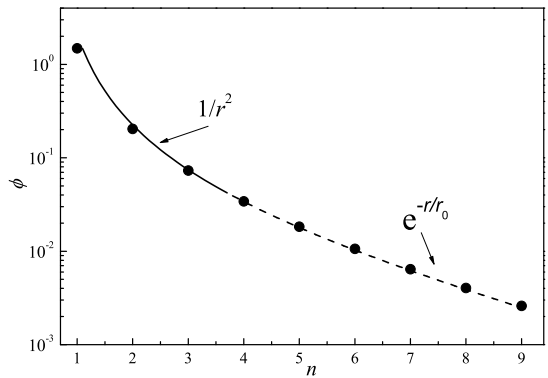}}
\caption{\label{FigPhiR}  Angles $\varphi_{n,n}$ (circles)
calculated by solving numerically Eqs.~\eqref{sys} at $t/J=50$,
$\varkappa=5 \times 10^{-2}$ ($r_0\simeq4.5$). The curve
corresponds to the function $\varphi(n-1/2,n-1/2)$, see
Eq.~\eqref{PhiAppr}. The initial angle $\varphi_0$ in
Eq.~\eqref{PhiAppr} is used as a fitting parameter. The range of
distances where the quantum fluctuations can play an important
role at $S=2$ is shown by the dashed line.}
\end{figure}
\begin{figure}
\centerline{\includegraphics[width=0.4\textwidth]{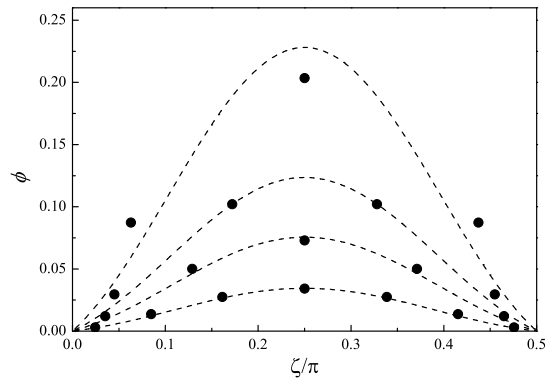}}
\caption{\label{FigPhiAng} $\varphi(x,y)$ vs $\zeta=\arctan(x/y)$
at $x+y=N$ for $N=3,\,4,\,5,\,7$ from the top to bottom calculated
according to Eq.~\eqref{PhiAppr} (dashed curves). Points
correspond to the numerical calculations of angles
$\varphi_{n,N-n+1}$ ($\zeta_n=\arctan[(n-1/2)/(N-n+1/2)]$).
Parameters of the model are $t/J=50$, $\varkappa=5\times
10^{-2}$.}
\end{figure}

The solution to Eq.~\eqref{LinEq} with boundary
condition~\eqref{Dirichlet} is
\begin{equation}\label{PhiAppr}
\varphi(\mathbf{r})=\frac{\varphi_0}{K_2(a/r_0)}K_2\left(\frac{r}{r_0}\right)\sin2\zeta,\,\,
r_0=\frac{1}{\sqrt{\varkappa}}\,,
\end{equation}
where $K_2(x)$ is the Macdonald function. Within the range
$|\mathbf{r}|<r_0$, $\varphi(\mathbf{r})$ behaves as $a^2/r^2$,
whereas at large distances, it decreases exponentially
$\varphi(\mathbf{r})\propto\exp(-r/r_0)$. The function
$\varphi(\mathbf{r})$ at $\zeta=\pi/4$ and the numerical results
for $\varphi_{n,n}$ are plotted in Fig.~\ref{FigPhiR}. In order to
compare the angular dependence of the approximate solution with
the numerical results, we plot in Fig.~\ref{FigPhiAng} the values
of $\varphi(x,y)$ at lines $x+y=N$ depending on the angle
$\zeta=\arctan(x/y)$ at different $N$. The values of
$\varphi_{n,N-n+1}$ are also shown in this figure. We see from
these figures that, except for the small region near the magnetic
polaron, the continuous function Eq.~\eqref{PhiAppr} is a good
interpolation of the function $\varphi_{n_x,n_y}$ at the discrete
lattice. Note that at small $r$, the continuum approximation fails
because of the large derivatives of $\varphi(\mathbf{r})$. In
addition, we cannot neglect the nonlinearity of differential
equation~\eqref{SinGordon} in this region.

\begin{figure}
\centerline{\includegraphics[width=0.38\textwidth]{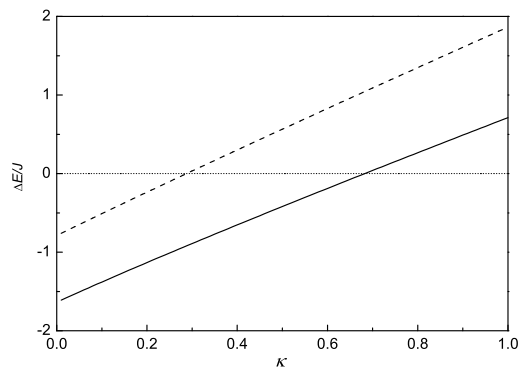}}
\caption{\label{FigDeltaE} The energy difference $\Delta E$
between 'coated' and 'bare' ferrons vs $\varkappa=2K/J$ calculated
at $t/J=50$. Solid curve corresponds to the numerical
calculations, whereas dashed curve is calculated using
Eq.~\eqref{EpAppr}.}
\end{figure}

Let us calculate now the energy of 'coated' magnetic polaron,
$E_{\text{p}}$, in comparison to the 'bare' one
$E_{\text{p}}^{0}=-2t+16J$. The energy of the system is given by
Eq.~\eqref{Etot}, where $\theta_{\mathbf{n}}=\pi/2$ and
$\varphi_{\mathbf{n}}$ is a solution to Eq.~\eqref{sys}. In the
continuum approximation, we expand $\cos\nu_{\mathbf{nm}}$ and
$\cos^2\varphi_{\mathbf{n}}$ in the Taylor series up to the second
order outside the ferron core. Changing the summation by
integration over $\mathbf{r}$, we find for the last two terms in
Eq.~\eqref{Etot}
\begin{equation}\label{Em}
8J\left(1+\frac{\varkappa}{4}\right)\sin^2\varphi_0%
+\frac{J}{2}\!\!\!\int\limits_{|\mathbf{r}|\geq
a}\!\!\!d^2r\left[({\bm\nabla}\varphi)^2+\varkappa\varphi^2\right]\,.
\end{equation}
In Eq.\eqref{Em}, the first term comes from the summation over
spins in the core of the magnetic polaron. Substituting
solution~\eqref{PhiAppr} into Eq.~\eqref{Em} and performing the
integration, we find
\begin{equation}\label{EpAppr}
E_{\text{p}}=8J\left(1+\frac{\varkappa}{4}\right)\sin^2\varphi_0+
\frac{J\varphi_0^2}{2}I(a\sqrt{\varkappa})-2t\sin\varphi_0\,,
\end{equation}
where
\begin{equation}
I(\rho_0)=2\pi\left[1+\frac{\rho_0K_1(\rho_0)}{2K_2(\rho_0)}\right]\,.
\end{equation}
The initial angle $\varphi_0$ is found by minimization of
energy~\eqref{EpAppr}. This value turns out to be slightly
different from that found numerically. The dependence of $\Delta
E=E_{\text{p}}-E_{\text{p}}^{0}$ on $\varkappa$ is shown in
Fig.~\ref{FigDeltaE}. We see from this figure that $\Delta E<0$ at
not too high values of anisotropy constant $K$. So, the magnetic
polaron with extended spin distortions can be more favorable than
the 'bare' one. Note also that formula~\eqref{EpAppr}
overestimates the energy of the 'coated' magnetic polaron.

\section{Magnetic polaron in 3D lattice}\label{sec3D}

In this section, we calculate the magnetic structure of the
'coated' magnetic polaron in the 3D case using the approach
described above. In the $V\to\infty$ limit, a magnetic polaron
includes $8$ sites, that is, $\Psi_{\mathbf{n}}\neq0$ only for
$\mathbf{n}_1=\{1,\,1,\,1\}$, $\mathbf{n}_2=\{1,\,0,\,1\}$,
$\mathbf{n}_3=\{0,\,0,\,1\}$, $\mathbf{n}_4=\{0,\,1,\,1\}$,
$\mathbf{n}_5=\{0,\,1,\,0\}$, $\mathbf{n}_6=\{1,\,1,\,0\}$,
$\mathbf{n}_7=\{1,\,0,\,0\}$, $\mathbf{n}_8=\{0,\,0,\,0\}$. The
electron energy is found then by the diagonalization of the
corresponding $8\times8$ matrix. As in previous section, we assume
that all spins in the lattice lie in the $xy$ plane. The symmetry
conditions then become
\begin{equation}\label{SymCond3D}
\left\{\begin{array}{rcl}
\varphi_1=\varphi_3=\varphi_5=\varphi_7&=&\varphi_0,\\
\varphi_2=\varphi_4=\varphi_6=\varphi_8&=&-\varphi_0\,.
\end{array}\right.
\end{equation}

Minimizing total energy~\eqref{Etot} with respect to
$\varphi_{\mathbf{n}}$, we get a system of equations similar to
Eqs.~\eqref{sys} with the only difference that the factor at the
sum in the right-hand side is equal now to $3t/8J$ and
${\bm\Delta}$ takes values $\{\pm1,\,0,\,0\}$, $\{0,\,\pm1,\,0\}$,
and $\{0,\,0,\,\pm1\}$. We solve this system of equations
numerically for the cluster with $30\times30\times30$ sites.

In the continuum approximation, the magnetic structure is found in
the same manner as in the 2D case. We consider a spherical
magnetic polaron of radius $a=\sqrt{3}/2$. The function
$\varphi(\mathbf{r})$ satisfies Eq.~\eqref{LinEq} with the
boundary conditions at $|\mathbf{r}|=a$
\begin{equation}\label{Dirichlet3D}
\varphi(\mathbf{r})|_{r=a}=\tilde{\varphi}(\vartheta,\zeta)\,,
\end{equation}
where $r$, $\vartheta$, and $\zeta$ are spherical coordinates and
the function $\tilde{\varphi}(\vartheta,\zeta)$ should satisfy the
symmetry conditions~\eqref{SymCond3D} at angles $\vartheta_i$,
$\zeta_i$ corresponding to the directions of
$\mathbf{n}_i-\mathbf{n}_0$ ($i=1\dots8$). Expanding
$\tilde{\varphi}(\vartheta,\zeta)$ in series of spherical
functions $\text{Y}_{lm}(\vartheta,\zeta)$ and retaining only the
terms with the smallest $l$, one finds
\begin{eqnarray}
\tilde{\varphi}(\vartheta,\zeta)&=&\varphi_0\sqrt{\frac{36\pi}{35}}%
\left[\text{Y}_{3,+2}(\vartheta,\zeta)-\text{Y}_{3,-2}(\vartheta,\zeta)\right]\nonumber\\
&=&\varphi_0\sqrt{\frac{27}{2}}\sin^2\vartheta\cos\vartheta\sin2\zeta\,.
\end{eqnarray}
The solution to Eq.~\eqref{LinEq} with the boundary
conditions~\eqref{Dirichlet3D} is
\begin{equation}\label{PhiAppr3D}
\varphi(x,y,z)=\varphi_0\frac{R_3(r/r_0)}{R_3(a/r_0)}\frac{3\sqrt{3}\,xyz}{r^3}\,,
\end{equation}
where again $r_0=1/\sqrt{\varkappa}$, and
\begin{equation}\label{R3}
R_3(\rho)=\left(1+\rho+\frac{2\rho^2}{5}+\frac{\rho^3}{15}\right)%
\displaystyle\frac{\text{e}^{-\rho}}{\rho^4}\,.
\end{equation}

\begin{figure}
\centerline{\includegraphics[width=0.45\textwidth]{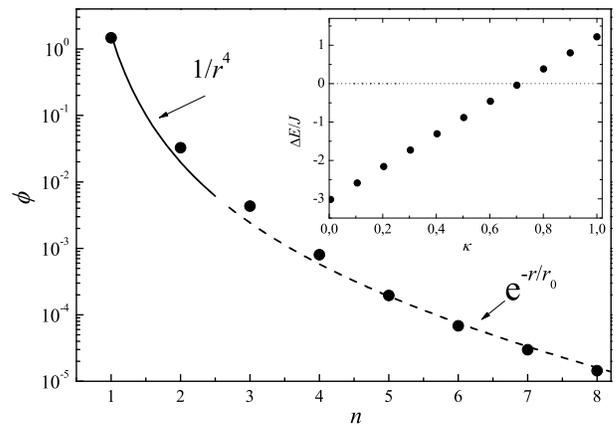}}
\caption{\label{FigPhiR3D}  The dependence of
$\varphi(\mathbf{r})$ on the distance $r$ in the direction
$\{1,\,1,\,1\}$ calculated according to Eq.~\eqref{PhiAppr3D}
(solid and dashed curve). Circles represent the numerical results
for $\varphi_{n,n,n}$ at $t/J=100$, $\varkappa=5 \times 10^{-2}$.
The initial angle $\varphi_0$ in Eq.~\eqref{PhiAppr3D} is used as
a fitting parameter. The range of distances where the quantum
fluctuations can play an important role at $S=2$ is shown by the
dashed line. In the inset, the dependence of $\Delta E$ on
$\varkappa$ is shown. The energy difference $\Delta E$ is
calculated numerically at $t/J=100$.}
\end{figure}

It follows from Eqs.~\eqref{PhiAppr3D} and~\eqref{R3} that
$\varphi(\mathbf{r})$ decreases at $r<r_0$ as $a^4/r^4$. At
distances $r$ exceeding characteristic radius $r_0$, the spin
distortions decay exponentially,
$\varphi(\mathbf{r})\propto\exp(-r/r_0)$ as in the 2D case. The
dependence of $\varphi(\mathbf{r})$ on distance $r$ in the
$\mathbf{n}_1=\{1,\,1,\,1\}$ direction is shown in
Fig.~\ref{FigPhiR3D}. The numerical results for $\varphi_{n,n,n}$
are also presented in this figure.

The energy of the magnetic polaron, $E_{\text{p}}$, is found in
the same way as in the previous section. Note that in the 3D case,
the energy of a 'bare' magnetic polaron is
$E_{\text{p}}^{0}=-3t+48J$. The dependence of $\Delta
E=E_{\text{p}}-E_{\text{p}}^{0}$ on $\varkappa$ is shown in the
inset to Fig.~\ref{FigPhiR3D}. We see that in the 3D case, the
situation is similar to that considered in the previous section.
At not too high anisotropy constant, the 'coated' magnetic polaron
is favorable in energy in comparison to the 'bare' one.

Note, however, that the two-sublattice AFM state considered here
is not eigenstate of the system, and quantum fluctuations play an
important role. Indeed, the fluctuation magnitude is about $\delta
S/S=0.078/S$ for the isotropic 3D Heisenberg model~\cite{Kittel}.
This value becomes of the order of the angle $\phi$ at $r=2$
(second coordination sphere). However, we consider the situation
with uniaxial anisotropy, which drastically reduces quantum
fluctuations stabilizing the two-sublattice AFM state. For
example, for manganites, the characteristic value of the
anisotropy parameter $K=0.05J$ (see Ref.~\onlinecite{Nikiforov}).
At such anisotropy, the fluctuation magnitude decreases by about
30\%. This means that we can use our results up to $r=3$. At the
concentrations of the impurities higher than 1\% the distance
between impurities is less than 6 lattice constants and the
behavior of the distortions at $r>3$ is not of real physical
importance. The similar arguments are valid also for the 2D case.
Here, we can find that $\delta S/S=0.092/S$ at $K=0.05J$ (note
that the magnetic anisotropy stabilizes the long-range magnetic
order in the 2D Heisenberg model). In 2D, the distortions of AFM
order decay slower than in the 3D case. Thus, the quantum
fluctuations become comparable to the angle $\phi$ at $r\simeq4$.
The range of distances where the quantum fluctuations can play an
important role at $S=2$ is shown by the dashed line in
Figs.~\ref{FigPhiR} and \ref{FigPhiR3D}.

\section{Conclusions}

We studied the structure of bound magnetic polarons in 2D and 3D
antiferromagnets doped by nonmagnetic donor impurities. The doping
concentration was assumed to be small enough, and the system is
far from the insulator-metal transition. It was also assumed that
the conduction electron is bound by electrostatic potential of
impurity with the localization length $a$ close to the lattice
constant $d$. We found two possible types of magnetic polarons,
'bare' and 'coated'. At not too high constant of magnetic
anisotropy $K\ll J$, the 'coated' polarons correspond to a lower
energy. Such a 'coated' ferron consists of a ferromagnetic core of
the size $a$ and extended spin distortions of the AFM matrix
around it. The characteristic radius of distortions is
$r_0=d\sqrt{J/2K}$. Within the $a<r<r_0$ range, these spin
distortions behave as $a^2/r^2$ in the 2D case and as $a^4/r^4$ in
the 3D case. At larger distances, they decay exponentially as
$\exp(-r/r_0)$. Note that in real systems, the constant of
magnetic anisotropy $K$ is small enough in comparison to exchange
integral $J$ and the characteristic radius $r_0$ can be rather
large.

The calculations have been performed in the case of zero
temperature $T=0$. At finite temperatures, the regular picture of
spin distortions could be, in principle, destroyed, owing to the
spin-wave excitations. However, in the case of uniaxial
anisotropy, the spin-wave spectrum has a gap of the order of
$\sqrt{2KJ}$. So, the characteristic temperature, below which our
results (in particular, those presented in Figs.~\ref{FigPhiR} and
\ref{FigPhiR3D}) are applicable, is of the same order. For
example, in manganites, a typical value of this temperature is
5\,K.

Note that in the 1D case, the 'coated' ferron was found to be a
metastable object~\cite{Fer1D} in contrast to the 2D and 3D cases
considered here. The cause of such a difference is the following.
In the 1D case, the 'bare' magnetic polaron has zero surface
energy since it does not disturb the AFM ordering at its border.
For $D>1$, the surface energy is nonzero due to the geometry of
the problem. For the 'coated' magnetic polaron, the surface energy
is nonzero at any dimension. However, the extended 'coat' reduces
its value.

Our results were obtained in the limit of strong electron-impurity
coupling, $V\to\infty$. In this case, the wave function
$\Psi_{\mathbf{n}}$ of the conduction electron is nonzero only at
the sites nearest to the impurity, and one can calculate the
electron energy $E_{\text{el}}$ exactly. At finite $V$, the wave
function extends over larger distances and one should calculate
the magnetic structure simultaneously with $\Psi_{\mathbf{n}}$.
Note, that even at $V=0$, the magnetic polaron is a stable object
in a wide range of parameters of the model. This is the so called
self-trapped magnetic polaron, which exists due to trapping of a
charge carrier in the potential well of ferromagnetically oriented
local spins~\cite{Nag67}. The radius of such a polaron was
estimated as $a\sim d(t/J)^{1/5}$ and at large $t/J$ it can be of
the order of several lattice constants~\cite{Kak,Kkm}.

Let us find the range of validity for the assumption $V\to\infty$
used in this paper. For this purpose, we rewrite the electron
energy $E_{\text{el}}$ in terms of the wave function. Performing
the transformation of angles described in Section~\ref{Mstruct},
and using the substitution
$\Psi_{\mathbf{n}}=\text{e}^{i\varphi_{\mathbf{n}}/2}\psi_{\mathbf{n}}$,
we can write at $\theta_{\mathbf{n}}=\pi/2$
$$
E_{\text{el}}=t\sum_{\langle\mathbf{nm}\rangle}%
\sin\left(\frac{\varphi_{\mathbf{n}}-\varphi_{\mathbf{m}}}{2}\right)%
\text{Im}\left(\psi_{\mathbf{n}}\psi_{\mathbf{m}}^{*}\right)%
-V\sum_{\mathbf{n}}\frac{\left|\psi_{\mathbf{n}}\right|^2}{|\mathbf{n-n}_0|}.
$$
Minimizing the total energy with respect to angles
$\varphi_{\mathbf{n}}$, we obtain the set of equations similar to
Eqs.~\eqref{sys} with the right-hand side in the form
\begin{equation}\label{rhs}
\frac{t}{2J}\sum_{\bm{\Delta}}\cos\left(\frac{\varphi_{\mathbf{n}+\bm{\Delta}}-\varphi_{\mathbf{n}}}{2}\right)%
\text{Im}\left(\psi_{\mathbf{n}}\psi_{\mathbf{n}+\bm{\Delta}}^{*}\right)\,.
\end{equation}
In the limit of strong electron-impurity coupling, we can assume
that the wave function is hydrogen-like, that is,
$|\psi_{\mathbf{n}}|\approx
N_{\text{p}}^{-1/2}\exp[-(d|\mathbf{n-n}_0|-a)/a_{B}^{*}]$, where
$N_{\text{p}}$ is the number of sites inside the magnetic polaron
and $a_{B}^{*}$ is the effective Bohr radius. The latter one can
be estimated as $a_{B}^{*}=\hbar^2\epsilon/m^{*}e^2$, where
$\epsilon$ is the permittivity, and $m^{*}$ is the effective
electron mass, which can be found from the relation
$\hbar^2/2m^{*}=td^2$. Taking into account these estimations, we
find that one can neglect the right-hand side~\eqref{rhs} in
Eq.~\eqref{sys} for $\mathbf{n}\neq\mathbf{n}_i$ if
$$
\frac{t}{2JN_{\text{p}}}\text{e}^{-V/2t}\ll1\,.
$$
In the 3D case, we have $N_{\text{p}}=8$, and the approximation
$V\to\infty$ is valid for $V=e^2/\epsilon d$ down to
$V_c=2t\ln(t/16J)$. For realistic values of parameters $V$ is
usually of the order of $V_c$. So, it is of interest to perform a
more detailed analysis of the problem at finite values of $V$
similar to that discussed in Ref.~\onlinecite{Ivan2} for the 1D
case. Nevertheless, we hope that our results concerning the
extended spin distortions are valid also at $V<V_c$, and,
consequently, for magnetic polarons with larger sizes $a$. Indeed,
in the continuum approximation, the dependence of the magnetic
structure on $a$ is not so important. Moreover, the energy of the
magnetic polaron $E_{\text{p}}$ formally depends on $a$ only
through the term $J\varphi_0^2I(a\sqrt{\varkappa})/2\sim J$ coming
from the spatial integration (see Eq.~\eqref{EpAppr}). Since
$\varkappa\ll1$, the energy difference $\Delta E$ is negative up
to $a\sim r_0\gg d$.

In our paper, we studied the isolated ferrons. However, as it was
already mentioned above, the substantial overlap of the extended
distortions can arise even at very low doping (about 1\%). As a
result, the magnetic moments of the ferrons could be ordered. In
particular, according to Ref.~\onlinecite{Nag01}, in such a
situation, there can exist a long-range ferromagnetic order with
the magnetic moment far from the saturation value.

\section*{Acknowledgments}

The authors are grateful to  A.\,V.~Klaptsov and I.~Gonz\'alez for
helpful discussions.

The work was supported by the Russian Foundation for Basic
Research, project No. 05-02-17600 and International Science and
Technology Center, grant No. G1335.

\end{document}